\documentclass[aps,prl,twocolumn,preprintnumbers,superscriptaddress]{revtex4-1}
\pdfoutput=1

\usepackage[usenames,dvipsnames,table]{xcolor}
\usepackage{graphicx,amsmath,amssymb,amsthm,multirow,array,bm,}
\usepackage[mathscr]{eucal}
\usepackage[bbgreekl]{mathbbol}
\usepackage{amsfonts}
\usepackage{hyperref}
\usepackage{tikz}
\usetikzlibrary{3d,calc}
\usepackage{ulem}

\newcommand{\beq}{\begin{equation}}
\newcommand{\eeq}{\end{equation}}
\newcommand{\II}{\mathrm{I}\hspace{-0.8pt}\mathrm{I}}
\def\<{\langle}
\def\>{\rangle}

\begin{document}

\title{A Constraint on Defect and Boundary Renormalization Group Flows}

\preprint{OUTP-15-19P}
\preprint{YITP-SB-15-33}

\author{Kristan Jensen}
\email{kristanj@insti.physics.sunysb.edu}
\affiliation{C.N. Yang Institute for Theoretical Physics, SUNY Stony Brook, Stony Brook, New York 11794, U.S.A.}

\author{Andy O'Bannon}
\email{obannon@physics.ox.ac.uk}
\affiliation{Rudolf Peierls Centre for Theoretical Physics, University of Oxford, 1 Keble Road, Oxford OX1~3NP, U.~K.}

\date{\today}

\begin{abstract}
A conformal field theory (CFT) in dimension $d\geq 3$ coupled to a planar, two-dimensional, conformal defect is characterized in part by a ``central charge'' $b$ that multiplies the Euler density in the defect's Weyl anomaly. For defect renormalization group flows, under which the bulk remains critical, we use reflection positivity to show that $b$ must decrease or remain constant from ultraviolet to infrared. Our result applies also to a CFT in $d=3$ flat space with a planar boundary.

\end{abstract}

\pacs{}

\maketitle

\textit{Introduction.}~Monotonicity theorems, such as Zamolodchikov's $c$-theorem~\cite{Zamolodchikov:1986gt}, are of fundamental importance in quantum field theory (QFT). They make precise the intuition that the number of degrees of freedom (DOF) should decrease under renormalization group (RG) flow. They therefore place stringent constraints on the low-energy physics of QFTs. For example, they can eliminate the possibility of RG limit cycles, and can eliminate potential low-energy dualities between QFTs (see \textit{e.g.}~\cite{Grover:2012sp}).

An ideal monotonicity theorem consists of six constraints on an observable $X$, treated as a function over the space of couplings in the QFT~\footnote{A pre-condition for $X$ to count DOF is that $X$ is independent of exactly marginal couplings.}:
\begin{enumerate}
\item The value of $X$ at the ultra-violet (UV) fixed point is greater than or equal to its value at the infra-red (IR) fixed point: $X_{\textrm{UV}}\geq X_{\textrm{IR}}$ (the ``weak'' form);
\item $X$ strictly decreases or remains constant along the RG flow (the ``strong form'');
\item $X$ decreases along a gradient along the RG flow (``strongest form'');
\item $X$ is stationary at fixed points (and nowhere else);
\item $X$ is bounded from below;
\item $X$ counts only non-topological DOF.
\end{enumerate}

These are listed roughly in decreasing order of importance: 1 is essential, 2 and 3 are highly desirable, and 4 through 6 are appealing but expendable. Obviously $3$ implies $2$, and $2$ implies $1$. While $1$, $5$ and $6$ can be deduced from fixed points alone, $2$, $3$, and $4$ require an ``$X$-function'' defined everywhere along the RG flow.

Ideally, the derivation of a monotonicity theorem should be non-perturbative, relying only on generic properties of a ``healthy'' QFT. To date, the standard assumptions are that the QFT is renormalizable, local, and for Lorentzian QFTs, Poincar\'e-invariant and unitary, or for Euclidean QFTs, Euclidean-invariant and reflection-positive. The only other, more restrictive, assumption is that RG fixed points are conformal field theories (CFTs).

The gold standard remains Zamolodchikov's $c$-theorem, for QFTs in dimension $d=2$~\cite{Zamolodchikov:1986gt}. Zamolodchikov identified $X$ as a particular linear combination of two-point functions of the stress tensor and its trace, called the ``$c$-function,'' which at fixed points reduces to the central charge $c$. Zamolodchikov established constraints $2$ and $4$ using the assumptions above, and constraint $3$ within conformal perturbation theory, while reflection positivity implies $5$ and $c$'s definition implies $6$.

Zamolodchikov's arguments rely crucially on the special form of the stress tensor two-point function in $d=2$, and are thus difficult to generalize to $d>2$. Moreover, for a CFT in $d=2$, a single number, $c$, fixes the Virasoro algebra, Weyl anomaly, thermal entropy, and more. The same is not true for CFTs in $d>2$, raising the question of which $X$ to target for a proof.

For even $d>2$, Cardy targeted $a$, the coefficient of the Euler density in the Weyl anomaly~\cite{Cardy:1988cwa}. By definition, $a$ satisfies constraint $6$. In $d=4$, positivity of energy flux at spatial infinity~\cite{Hofman:2008ar} implies that $a$ satisfies constraint $5$. Moreover, in $d=4$ Jack and Osborn established a strong $a$-theorem valid to all orders in perturbation theory~\cite{Osborn:1989td,Jack:1990eb,Osborn:1991gm}, although their method, based on local Weyl consistency, is difficult to generalize to $d>4$~\cite{Osborn:2015rna}. Komargodski and Schwimmer provided a non-perturbative argument for the weak form in $d=4$, $a_{\textrm{UV}}\geq a_{\textrm{IR}}$~\cite{Komargodski:2011vj,Komargodski:2011xv} (see also~\cite{Luty:2012ww}). Their method, which uses an external scalar field to match UV and IR Weyl anomalies~\cite{Schwimmer:2010za}, is also difficult to generalize to $d>4$~\cite{Elvang:2012st}. Evidence for an $a$-theorem in $d=6$ appears in~\cite{Cordova:2015vwa,Cordova:2015fha,Heckman:2015axa}.

No Weyl anomaly exists in odd $d$, making these cases more challenging still. To date, the leading candidate for $X$ is the sphere ``free energy'' $F \equiv (-1)^{(d-1)/2}\ln Z_{\mathbb{S}^d}$, with $Z_{\mathbb{S}^d}$ the renormalized partition function of a Euclidean CFT on a sphere, $\mathbb{S}^d$~\cite{Myers:2010tj,Jafferis:2011zi}. In $d=1$, $\mathbb{S}^1$ is the ``thermal circle,'' so $F$ is minus the thermal free energy. Positivity of the heat capacity then immediately implies a strong $F$-theorem. In $d=3$, $F \neq 0$ in pure Chern-Simons theory~\cite{Witten:1988hf}, manifestly violating constraint $6$. Using a relation between $F$ and disk entanglement entropy (EE) at fixed points~\cite{Casini:2011kv}, Casini and Huerta established a strong $F$-theorem using strong subadditivity of EE~\cite{Casini:2012ei}. However, their $F$-function violates constraint $4$~\cite{Klebanov:2012va}. An alternative is mutual information, which obeys constraint $2$ and possibly $5$ and $6$, but violates $4$~\cite{Casini:2015woa}. For discussions about $F$-theorems in $d>3$, see for example~\cite{Jafferis:2012iv,Giombi:2014xxa}.

Another class of monotonicity theorems concern DOF at a boundary or defect. For example, consider a boundary CFT (BCFT), \textit{i.e.}\ a CFT on a space with a boundary, with conformally-invariant boundary conditions (BC). Under a boundary RG flow, triggered by a relevant operator at the boundary, the bulk remains critical, and the IR fixed point is again a BCFT. For such RG flows in $d=2$, Affleck and Ludwig proposed a monotonicity theorem for $\ln g \equiv - \ln Z_{\mathbb{HS}^2} + \frac{1}{2}\ln Z_{\mathbb{S}^2}$, with $Z_{\mathbb{HS}^2}$ the BCFT partition function on a hemisphere, $\mathbb{HS}^2$~\cite{Affleck:1991tk}. Friedan and Konechny established a strongest $g$-theorem using thermodynamic entropy~\cite{Friedan:2003yc}. The $g$-theorem applies also to point-like defects, via the folding trick. Conjectures for $g$-theorems in $d>2$ appear in~\cite{Nozaki:2012qd,Jensen:2013lxa,Estes:2014hka,Gaiotto:2014gha}.

In this Letter we establish a weak $g$-theorem for Euclidean BCFTs in $d=3$, and for Euclidean defect CFTs (DCFTs) in $d \geq 3$ with a two-dimensional planar defect. Our $X$ is $b$, the coefficient of the Euler density in the boundary or defect Weyl anomaly. Using the standard assumptions above, we establish $b_{\textrm{UV}} \geq b_{\textrm{IR}}$ for boundary or defect RG flows. Our argument is an adaptation of Komargodski's argument for the weak $c$-theorem~\cite{Komargodski:2011xv}. Ultimately, our ``$b$-theorem'' is equivalent to the conjectures of~\cite{Nozaki:2012qd,Jensen:2013lxa,Estes:2014hka} for two-dimensional defects or boundaries.

%%%%%%%%%%%%%%%%%%%%%%%%%%%%%%%%%%%%%%%%%%%%%%%%%%%%%%%%%%%%%%%%%%%%%%%%%%%%%%%%%%%%%%%%%%%%%%%

\textit{The Systems.} We begin with local, reflection-positive, parity-invariant Euclidean CFTs in $d\geq3$. Ultimately we are interested in these CFTs in flat space, but to study their Weyl anomalies we will put them in curved space, unless stated otherwise. We thus introduce an external metric $g_{\mu\nu}$\footnote{In order to avoid a curvature-induced singularity in the generating functional, \textit{i.e.}\ a phase transition, we consider only weakly-curved perturbations of the flat metric, unless stated otherwise.}. The CFT's generating functional of renormalized, connected correlators, $W \equiv - \ln Z[g_{\mu\nu}]$, with $Z[g_{\mu\nu}]$ the renormalized partition function, is invariant under coordinate reparameterizations and Weyl transformations, $g_{\mu\nu}\to e^{2\omega} g_{\mu\nu}$ (with $\omega$ a real function of space), up to the Weyl anomaly. These invariances imply that the flat-space theory is invariant under the action of the conformal algebra, $\mathfrak{so}(d+1,1)$, generated by infinitesimal rotations, translations, dilatations, and special conformal transformations.

Next we introduce a two-dimensional defect. For example, we can impose BC on CFT fields along a two-dimensional subspace, or introduce fields localized there, with or without couplings to the bulk CFT. Although ultimately we are interested in flat-space CFTs with planar defects, we will put them in curved space, and keep the defect's position arbitrary, unless stated otherwise. We assume the defect preserves locality, reflection positivity~\footnote{For a BCFT on a half-plane, the state-operator correspondence implies that reflection positivity of the Euclidean theory is equivalent to unitarity of the Lorentzian theory on a hemisphere. However, for more general boundary or defect QFTs, whether reflection positivity is equivalent to unitarity is an open question. The proofs of equivalence assume the full Euclidean symmetry~\cite{Osterwalder:1973dx,Osterwalder:1974tc}. A boundary or defect breaks that symmetry to a subgroup, hence strictly speaking those proofs are no longer valid. We are not aware of extensions of those proofs to cases with boundaries or defects. Nevertheless, we suspect only some ``pathology,'' for example non-locality, could spoil the equivalence. If the equivalence is valid, then all of our results extend to Lorentzian theories with unitary defect RG flows, via Wick rotation.}, parity, and reparameterization and Weyl invariances, up to a Weyl anomaly. The resulting theory is a DCFT.  Reparameterization and Weyl invariance imply that the flat-space DCFTs are invariant under the action of the $\mathfrak{so}(d-1,1)\times \mathfrak{so}(d-2)$ subalgebra of $\mathfrak{so}(d+1,1)$ that preserves the planar defect~\footnote{Although the defect is two-dimensional, this ``defect conformal algebra'' does not contain Virasoro factors.}.

Another option, special to $d=3$, is that the bulk CFT changes across the defect. Indeed, a BCFT can be viewed as a DCFT with an ``empty'' CFT on one side of the defect. Our results will thus apply to BCFTs, but we will only explicitly discuss DCFTs, unless stated otherwise.

We are interested in defect RG flows in flat space, meaning flows triggered by a relevant operator at the defect, whose endpoints are flat-space DCFTs with planar defects. For example, consider a DCFT described by a local Lagrangian $\mathcal{L}_{\textrm{DCFT}} = \mathcal{L}_{\textrm{CFT}} +  \delta^{d-2}\mathcal{L}_{\textrm{defect}}$, with $\mathcal{L}_{\textrm{CFT}}$ the bulk CFT's Lagrangian, $\delta^{d-2}$ a Dirac delta function which restricts to the defect, and $\mathcal{L}_{\textrm{defect}}$ representing all defect terms. We trigger a defect RG flow by deforming $\mathcal{L}_{\textrm{DCFT}} \to \mathcal{L}_{\textrm{DCFT}} + \delta^{d-2}\lambda \, \mathcal{O}$, with $\mathcal{O}$ a dimension $\Delta_{\textrm{UV}} < 2$ parity-invariant scalar operator, and $\lambda$ a dimensionful coupling constant. Such an $\mathcal{O}$ may be built out of defect fields alone, bulk operators evaluated at the defect, or both. For example, we can give masses to defect fields, or change the BC on bulk fields.

Returning to curved space and defects of arbitrary position, let $x^{\mu}$ and $\sigma^a$ ($a=1,2$) be bulk and defect coordinates. Embedding functions $X^{\mu}(\sigma^a)$ then describe the defect's position. The defect's induced metric, $\hat{g}_{ab} \equiv g_{\mu\nu} \partial_a X^{\mu}\partial_b X^{\nu}$, describes the defect's intrinsic curvature. The bulk covariant derivative, $\nabla_{\mu}$, induces a defect covariant derivative, $\hat{\nabla}_a$. The second fundamental form, $\II^{\mu}_{ab} \equiv \hat{\nabla}_a \partial_b X^{\mu}$, describes the defect's extrinsic curvature. More details about the defect's geometry appear in Appendix A.

A key ingredient for us will be the stress tensor, $T^{\mu\nu}$. We define renormalized, connected correlators of $T^{\mu\nu}$, and the ``displacement operator,'' $D_{\mu}$, as follows. The renormalized partition function $Z$ is a functional of $g_{\mu\nu}$, $X^{\mu}$, and the set of all marginal or relevant couplings, $\{\lambda\}$. We define one-point functions $\langle T^{\mu\nu}\rangle$ and $ \langle D_{\mu} \rangle$ from the variation of $W \equiv -\ln Z[g_{\mu\nu},X^{\mu},\{\lambda\}]$ with respect to $g_{\mu\nu}$ and $X^{\mu}$, respectively:
\begin{align}
\label{E:deltaW}
\delta W = &-\frac{1}{2}\int d^dx \sqrt{g}\, \delta g_{\mu\nu}\langle T_{\textrm{b}}^{\mu\nu} \rangle
\\
\nonumber
& \quad -\int d^2\sigma \sqrt{\hat{g}} \left[ \frac{1}{2}\delta g_{\mu\nu} \langle T_{\textrm{d}}^{\mu\nu}\rangle +  \delta X^{\mu} \langle D_{\mu}\rangle  + \hdots\right],
\end{align}
where $g$ and $\hat{g}$ are the determinants of $g_{\mu\nu}$ and $\hat{g}_{ab}$, respectively, and $\ldots$ indicates possible terms involving derivatives of $\delta g_{\mu\nu}$ normal to the defect. Re-writing the defect's volume as $\int d^2\sigma \sqrt{\hat{g}}=\int d^dx \sqrt{g} \, \delta^{d-2}$, we see from~\eqref{E:deltaW} that $\langle T^{\mu\nu} \rangle$ receives distinct bulk and defect contributions (hence the subscripts),
\beq
\langle T^{\mu\nu}\rangle = \langle T^{\mu\nu}_{\textrm{b}}\rangle + \delta^{d-2} \langle T^{\mu\nu}_{\textrm{d}}\rangle + \hdots,
\eeq
where $\ldots$ indicates terms involving normal derivatives of $\delta^{d-2}$, coming from the $\ldots$ in~\eqref{E:deltaW}. Higher-order variations of $W$ give higher-point correlators, in the usual way.

Reparameterization invariance leads to Ward identities relating $\langle T^{\mu\nu}\rangle $ and $\langle D_{\mu}\rangle$, which we present in Appendix B (specifically~\eqref{E:reparWard}). However, we only need one fact about the reparameterization Ward identities: the defect stress tensor $T^{\mu\nu}_{\textrm{d}}$ is not conserved. Energy and momentum can flow between bulk and defect, violating conservation of $T^{\mu\nu}_{\textrm{d}}$. As a result, we cannot simply copy Zamolodchikov's derivation of the $c$-theorem, which relies crucially on conservation of the two-dimensional stress tensor. That is why we turn instead to Komargodski and Schwimmer's method~\cite{Komargodski:2011vj,Komargodski:2011xv}, based on Weyl anomaly matching~\cite{Schwimmer:2010za}.

%%%%%%%%%%%%%%%%%%%%%%%%%%%%%%%%%%%%%%%%%%%%%%%%%%%%%%%%%%%%%%%%%%%%%%%%%%%%%%%%%%%%%%%%%%%%%%%

\textit{Weyl Anomaly.} CFTs are Weyl-invariant only up to a potential anomaly. That is, $W$ may change under an infinitesimal Weyl variation, $\delta_{\omega} g_{\mu\nu}=2\omega g_{\mu\nu}$, $\delta_{\omega} X^{\mu}=0$:
\beq
\label{E:WeylAnomaly}
\delta_{\omega} W = -\int d^dx \sqrt{g}\, \omega \, \mathcal{A},
\eeq
where the local function $\mathcal{A}$ is built out of external fields, such as $g_{\mu\nu}$. Indeed, we will only consider contributions to $\mathcal{A}$ built from $g_{\mu\nu}$ alone~\footnote{Besides terms built from $g_{\mu\nu}$ alone, the only other terms that can appear in the anomaly $\mathcal{A}$ are built from $g_{\mu\nu}$ and sources conjugate to operators of appropriate dimension (see for example~\cite{Petkou:1999fv}). However, such terms are always Weyl-covariant (type B in the classification of~\cite{Deser:1993yx}), and hence will not crucially affect our arguments, and will not alter our result. We may therefore safely ignore them.}. Comparing~\eqref{E:WeylAnomaly} with~\eqref{E:deltaW} leads to the Weyl Ward identity, $\langle T^{\mu}_{~\mu}\rangle = \mathcal{A}$. The general form of $\mathcal{A}$ can be determined by solving the Wess-Zumino (WZ) consistency condition~\cite{Wess:1971yu}, which comes from demanding that two successive Weyl transformations of $W$ commute (the Weyl group is Abelian). For a CFT in even $d$, solving the WZ consistency condition gives~\cite{Deser:1993yx}
\beq
\label{E:bulkAnomaly}
\mathcal{A} = (-1)^{\frac{d}{2}+1}\frac{4a}{d!\text{vol}(\mathbb{S}^d)}E_d + \sum_I c_I W_I,
\eeq
with $E_d$ the Euler density and the $W_I$ the Weyl-covariant scalars of weight $-d$. WZ consistency allows total derivatives in~\eqref{E:bulkAnomaly}, which we eliminated using local counterterms. WZ consistency leaves undetermined the ``central charges'' $a$ and the $c_I$. For odd $d$, $\mathcal{A}=0$~\cite{Deser:1993yx}.

In a DCFT, $\mathcal{A}$ receives distinct bulk and defect contributions, $\mathcal{A} = \mathcal{A}_{\textrm{b}} + \delta^{d-2} \mathcal{A}_{\textrm{d}}$, where the bulk term $\mathcal{A}_{\textrm{b}}$ takes the form for $\mathcal{A}$ in a CFT, described above. To our knowledge, for the defect term, $\mathcal{A}_{\textrm{d}}$, the WZ consistency condition has been solved in only two cases: for a point-like defect in $d=2$~\cite{Polchinski:1998rq} and for a two-dimensional defect in $d\geq 3$~\cite{Henningson:1999xi,Schwimmer:2008yh} (sometimes called the ``Graham-Witten'' anomaly~\cite{Graham:1999pm}). We require the latter, which is, using local counterterms to cancel normal derivative terms~\cite{Henningson:1999xi,Schwimmer:2008yh},
\beq
\label{E:defectAnomaly}
\mathcal{A}_{\textrm{d}} = \frac{1}{24\pi}\left( b \, \hat{R}+ d_1 \, \mathring{\II}^{\mu}_{ab}\mathring{\II}_{\mu}^{ab} + d_2 \, W_{abcd} \hat{g}^{ac}\hat{g}^{bd}\right),
\eeq
with $\hat{R}$ the Ricci scalar of $\hat{g}_{ab}$, $\mathring{\II}^{\mu}_{ab}$ the traceless part of $\II^{\mu}_{ab}$, and $W_{abcd}$ the pullback of the bulk Weyl tensor. WZ consistency leaves undetermined the ``defect central charges'' $b$, $d_1$, and $d_2$. (The Weyl tensor vanishes identically in $d=3$, so $d_2$ exists only for $d \geq 4$.)

Under a Weyl transformation, $\sqrt{\hat{g}} \, \hat{R}$ transforms by a total derivative (that term is type A in the classification of~\cite{Deser:1993yx}), while $\sqrt{\hat{g}}\,\mathring{\II}^{\mu}_{ab}\mathring{\II}_{\mu}^{ab}$ and $\sqrt{\hat{g}} \, W_{ab}^{~~ab}$ are each Weyl-invariant (type B). Our $b$ is thus analogous to $a$, which obeys the $c$- or $a$-theorem in $d=2$ or $4$, respectively, while $d_1$ and $d_2$ are analogous to the $c_I$.

%%%%%%%%%%%%%%%%%%%%%%%%%%%%%%%%%%%%%%%%%%%%%%%%%%%%%%%%%%%%%%%%%%%%%%%%%%%%%%%%%%%%%%%%%%%%%%%

\textit{Monotonicity of $b$.} We will now argue that $b_{\textrm{UV}} \geq b_{\textrm{IR}}$ for defect RG flows, using Komargodski and Schwimmer's method~\cite{Komargodski:2011vj}. In particular, we will closely follow Komargodski's argument for the weak $c$-theorem~\cite{Komargodski:2011xv}.

Explicit breaking of Weyl invariance implies $\langle T^{\mu}_{~\mu} \rangle \neq \mathcal{A}$. In flat space with a planar defect, $\mathcal{A}=0$, so explicit breaking of Weyl invariance implies $\langle T^{\mu}_{~\mu} \rangle \neq 0$. For a defect RG flow, that occurs only at the defect: $\langle (T_{\textrm{d}})^{\mu}_{~\mu} \rangle \neq 0$, while $\langle (T_{\textrm{b}})^{\mu}_{~\mu} \rangle=0$, up to contact terms at the defect~\cite{Friedan:2003yc}. In curved space of even $d$ and/or for a curved defect, generically $\mathcal{A}\neq 0$. In that case, for a defect RG flow, explicit breaking of Weyl invariance only at the defect may lead to different defect central charges in the UV and IR, while bulk central charges will remain unchanged: $\mathcal{A}_{\textrm{d}}^{\textrm{UV}}\neq\mathcal{A}_{\textrm{d}}^{\textrm{IR}}$ while $\mathcal{A}_{\textrm{b}}^{\textrm{UV}}=\mathcal{A}_{\textrm{b}}^{\textrm{IR}}$.

However, we can undo explicit breaking of Weyl invariance by treating every relevant coupling $\lambda$ as a ``spurion.'' That is, we promote $\lambda$ to a function of defect coordinates, $\lambda \to \lambda(\sigma^a)$, and then endow $\lambda(\sigma^a)$ with a non-trivial Weyl transformation to restore Weyl invariance, up to the anomaly, leading to a modified Weyl Ward identity. Concretely, for a DCFT with a Lagrangian deformed as $\mathcal{L}_{\textrm{DCFT}} \to \mathcal{L}_{\textrm{DCFT}} + \delta^{d-2}\lambda \mathcal{O}$, as described above, we take $\lambda\to\lambda(\sigma^a)$, and under $g_{\mu\nu} \to e^{2 \omega} g_{\mu\nu}$ we demand $\lambda(\sigma^a) \to e^{(\Delta_{\textrm{UV}}-2)\omega}\lambda(\sigma^a)$. Following~\cite{Komargodski:2011vj,Komargodski:2011xv}, we will implement such a spurionic Weyl invariance using a non-dynamical, external scalar field, $\tau$~\footnote{When Weyl invariance is broken spontaneously, $\tau$ is the associated dynamical Goldstone boson. In that context, $\tau$ is called the ``dilaton.'' We avoid that name because the Coleman-Mermin-Wagner theorem forbids spontaneous breaking of Weyl invariance along a two-dimensional defect.}.
 Specifically, we re-define $\lambda \to \lambda' e^{(\Delta_{\textrm{UV}}-2)\tau}$, and under $g_{\mu\nu} \to e^{2 \omega} g_{\mu\nu}$ we demand $\tau \to \tau + \omega$ and $\lambda' \to \lambda'$.

The renormalized partition function $Z$ is now a functional of $g_{\mu\nu}$, $X^{\mu}$, and $\tau$, as well as the set of couplings $\{\lambda'\}$. We define $\mathcal{T}$ as the operator conjugate to $\tau$,
\beq
\langle \mathcal{T}\rangle \equiv \frac{1}{\sqrt{\hat{g}}}\frac{\delta W}{\delta \tau}.
\eeq
Under an infinitesimal Weyl variation, $\delta W$ takes the form in~\eqref{E:deltaW}, with $\delta_{\omega} g_{\mu\nu} = 2 \omega g_{\mu\nu}$, $\delta_{\omega} X^{\mu} = 0$, and now an ``extra'' term $\int d^2 \sigma \sqrt{\hat{g}} \langle \mathcal{T}\rangle \delta\tau$ with $\delta \tau = \omega$. From~\eqref{E:WeylAnomaly} we thus find
\beq
\label{E:newWeylWard}
\langle T^{\mu}_{~\mu} \rangle - \delta^{d-2}\langle \mathcal{T}\rangle = \mathcal{A},
\eeq
so that the Weyl Ward identity is unmodified in the bulk, $\langle (T_{\textrm{b}})^{\mu}_{~\mu} \rangle = \mathcal{A}_{\textrm{b}}$, but modified at the defect. In flat space with a planar defect, where $\mathcal{A}=0$,~\eqref{E:newWeylWard} says that $\langle \mathcal{T} \rangle$ cancels $\langle (T_{\textrm{d}})^{\mu}_{~\mu} \rangle \neq 0$ and any contact terms in $\langle (T_{\textrm{b}})^{\mu}_{~\mu} \rangle$, and thus restores Weyl invariance, as advertised ($\tau$ is a ``conformal compensator''). In curved space of even $d$ and/or with a curved defect, where generically $\mathcal{A}\neq0$,~\eqref{E:newWeylWard} says that $\langle \mathcal{T} \rangle$ acts to maintain $\mathcal{A}$'s UV value at all scales, including in particular the value at the defect. In other words, $\tau$ must account for the difference $\mathcal{A}_{\textrm{d}}^{\textrm{UV}}-\mathcal{A}_{\textrm{d}}^{\textrm{IR}}\neq0$. This is Weyl anomaly matching~\cite{Schwimmer:2010za}.

In flat space with a planar defect, the result $\langle T^{\mu}_{~\mu}\rangle=\delta^{d-2}\langle \mathcal{T}\rangle$ shows that $\tau$ becomes conjugate to $(T_{\textrm{d}})^{\mu}_{~\mu}$ plus contact terms in $(T_{\textrm{b}})^{\mu}_{~\mu}$. As a result, $\langle \mathcal{T}(\sigma)\mathcal{T}(0)\rangle$ has the same long-distance behavior as the two-point function of $T^{\mu}_{~\mu}$ in a $d=2$ flat-space QFT,
\beq
\label{E:IRTT}
\langle \mathcal{T}(\sigma)\mathcal{T}(0)\rangle  \propto \frac{1}{|\sigma|^{2\Delta_{\textrm{IR}}}},
\eeq
where $\Delta_{\textrm{IR}}>2$ is the dimension of the leading irrelevant deformation at the defect of the IR DCFT. The defect's planar symmetry and~\eqref{E:IRTT} together imply that the most general form for $\langle \mathcal{T}(\sigma)\mathcal{T}(0)\rangle$'s Fourier transform is, for small momentum $k$ along the defect,
\beq
\label{E:TT}
\langle \mathcal{T}(k)\mathcal{T}(-k)\rangle = \alpha_0 + \alpha_2 k^2 + \mathcal{O}(k^{2\Delta_{\textrm{IR}}-2}),
\eeq
where $\alpha_0$ and $\alpha_2$ are constants that can depend on $\{\lambda'\}$, and the $\mathcal{O}(k^{2\Delta_{\textrm{IR}}-2})$ terms arise from~\eqref{E:IRTT}. For small $k$ the ``soft'' $\mathcal{O}(k^{2\Delta_{\textrm{IR}}-2})$ terms are sub-leading compared to the contact terms $\alpha_1$ and $\alpha_2 k^2$, because $2\Delta_{\textrm{IR}}-2>2$. Similar statements apply for higher-point correlators of $\mathcal{T}$ with itself and with $T^{\mu\nu}_{\textrm{d}}$. The IR DCFT's effects on $\mathcal{T}$'s correlators are thus suppressed at small $k$, or equivalently, in the IR $\tau$ decouples from the IR DCFT.

That decoupling will persist to $g_{\mu\nu}\neq \delta_{\mu\nu}$, and will be explicit in the low-energy Wilsonian effective action:
\beq
\label{E:Seff}
S_{\textrm{eff}}= S_{\textrm{DCFT}}^{\textrm{IR}}+ S_{\tau}+ \mathcal{O}(\partial^{2\Delta_{\textrm{IR}}-2}),
\eeq
where $S_{\textrm{DCFT}}^{\textrm{IR}}$ is the IR DCFT's effective action, $S_{\tau}$ is $\tau$'s effective action, up to two derivatives, and $\mathcal{O}(\partial^{2\Delta_{\textrm{IR}}-2})$ represents $\tau$'s soft couplings to the IR DCFT. All terms in~\eqref{E:Seff} are functionals of $g_{\mu\nu}$, $X^{\mu}$, and $\tau$, except $S_{\textrm{DCFT}}^{\textrm{IR}}$, which does not depend on $\tau$ because of the decoupling.

Since $\tau$ has support only at the defect, $S_{\tau}$ consists of terms only at the defect. Under an infinitesimal Weyl variation, $\delta_{\omega} S_{\textrm{DCFT}}^{\textrm{IR}}$ produces the IR Weyl anomaly, $\mathcal{A}^{\textrm{IR}}=\mathcal{A}^{\textrm{UV}}_{\textrm{b}}+\mathcal{A}^{\textrm{IR}}_{\textrm{d}}$, so for Weyl anomaly matching $S_{\tau}$ must include WZ terms, $S_{\textrm{WZ}}$, such that $\delta_{\omega} S_{\textrm{WZ}}$ produces $\mathcal{A}_{\textrm{d}}^{\textrm{UV}}-\mathcal{A}_{\textrm{d}}^{\textrm{IR}}$. Together with locality and reparameterization invariance, that fixes $S_{\tau}$'s form (superscripts count derivatives of $\tau$)~\cite{Komargodski:2011xv}:
\begin{align}
\label{E:Stau}
S_{\tau} &\equiv S^{(0)} + S_{\textrm{WZ}}^{(0)}+S_\textrm{WZ}^{(2)},
\\
\nonumber
S^{(0)} & \equiv \int d^2\sigma\sqrt{\hat{g}} \left\{ -\frac{\beta_0}{4}e^{-2\tau} + \beta_1 \hat{R}+\beta_2 \mathring{\II}^2 + \beta_3 W_{ab}{}^{ab}\right\},
\\
\nonumber
S_{\textrm{WZ}}^{(0)} & \equiv -\frac{1}{24\pi}\int d^2\sigma \sqrt{\hat{g}} \, \tau \left\{\Delta b \hat{R} +  \Delta d_1 \mathring{\II}^2  +\Delta d_2 W_{ab}{}^{ab}\right\},
\\
\nonumber
S_{\textrm{WZ}}^{(2)}& \equiv \frac{\Delta b}{24\pi}\int d^2\sigma\sqrt{\hat{g}}\,\partial_a \tau \partial^a \tau,
\end{align}
where $\beta_0, \ldots, \beta_3$ are constants that can depend on $\{\lambda'\}$, while $\Delta b \equiv b_{\textrm{UV}}-b_{\textrm{IR}}$, and similarly for $\Delta d_1$ and $\Delta d_2$.

In~\eqref{E:Stau}, if we set $g_{\mu\nu}=\delta_{\mu\nu}$, Fourier transform, compute $\langle \mathcal{T}(k)\mathcal{T}(-k)\rangle$, and compare to~\eqref{E:TT}, then we find $\beta_0=\alpha_0$ and $\Delta b=-12 \pi \alpha_2$. The latter result provides a flat-space definition of $\Delta b$, and after a Fourier transform back to position space, implies a sum rule~\cite{Komargodski:2011xv}
\beq
\label{E:theSumRule}
b_{\textrm{UV}} - b_{\textrm{IR}} = 3\pi \int d^2\sigma |\sigma|^2 \langle \mathcal{T}(\sigma)\mathcal{T}(0)\rangle.
\eeq
The integral in~\eqref{E:theSumRule} is finite by power counting, plus no counterterms exist that can contribute to the right-hand-side of~\eqref{E:theSumRule}. Demanding reflection positivity in~\eqref{E:theSumRule}, $\langle \mathcal{T}(\sigma)\mathcal{T}(0)\rangle \geq 0$, thus leads to our main result,
\beq
\label{E:result}
b_{\textrm{UV}} \geq b_{\textrm{IR}}.
\eeq

For a marginally relevant deformation, $\langle \mathcal{T}(\sigma)\mathcal{T}(0)\rangle$ behaves at small $|\sigma|$ as $(\ln |\sigma|)/\sigma^4$. However, the integral in~\eqref{E:theSumRule} still converges, so again we find~\eqref{E:result}~\cite{Komargodski:2011xv,Luty:2012ww}.

%%%%%%%%%%%%%%%%%%%%%%%%%%%%%%%%%%%%%%%%%%%%%%%%%%%%%%%%%%%%%%%%%%%%%%%%%%%%%%%%%%%%%%%%%%%%%%%

\textit{Tests.} We test our result~\eqref{E:result} in four examples.

First is the free scalar BCFT in $d=3$, with a Neumann BC. A defect mass term triggers a defect RG flow to a Dirichlet BC. In Appendix C, we compute $b=1/16$ for the Neumann BC (correcting a result of~\cite{Nozaki:2012qd}) and $b=-1/16$ for the Dirichlet BC, so indeed $b_{\textrm{UV}} > b_{\textrm{IR}}$. The result $b<0$ for the Dirichlet BC raises the question of whether $b$ is bounded from below (constraint $5$).

Second is a DCFT in $d\geq3$ with a $d=2$ CFT of central charge $c$ added to the defect, but decoupled from all DCFT fields. That sends $b \to b + c$, but leaves unchanged $d_1$, $d_2$, and any bulk central charges. If we deform the $d=2$ CFT by a relevant scalar operator of the $d=2$ CFT, then the weak $c$-theorem implies $b_{\textrm{UV}} > b_{\textrm{IR}}$. Clearly $b$ can count DOF localized at the defect.

Third is a DCFT deformed by a weakly relevant defect operator: $\Delta_{\textrm{UV}} = 2- \varepsilon$ with $\varepsilon \ll 1$. The argument for the weak $c$-theorem based on perturbation theory in $\varepsilon$~\cite{Zamolodchikov:1986gt,Ludwig:1987gs}, trivially modified for a defect, gives $b_{\textrm{UV}} > b_{\textrm{IR}}$.

Fourth is the $\mathcal{N}=6$ supersymmetric (SUSY), strongly-coupled $U(N)_k\times U(N)_{-k}$ Chern-Simons matter theory~\cite{Aharony:2008ug} with $N$ and $N/k^5\gg1$, coupled to $N_f$ bi-fundamental hypermultiplet flavor fields at a two-dimensional defect, preserving $\mathcal{N}=(3,3)$ SUSY, with $N_f \ll N$~\cite{Ammon:2009wc}. That DCFT is holographically dual to $d=11$ supergravity on $d=4$ Anti-de Sitter space, $AdS_4$, times $\mathbb{S}^7/\mathbb{Z}_k$, with $N$ units of four-form flux, plus $N_f$ probe M5-branes along $AdS_3 \times \mathbb{S^3}/\mathbb{Z}_k$. Graham and Witten's holographic result~\cite{Graham:1999pm} gives $b=\frac{3}{2} N N_f$. A SUSY mass for $\Delta N_f$ of the hypermultiplets triggers a defect RG flow to the same DCFT, but now with $N_f - \Delta N_f$ hypermultiplets, hence $b_{\textrm{UV}} > b_{\textrm{IR}}$.

%%%%%%%%%%%%%%%%%%%%%%%%%%%%%%%%%%%%%%%%%%%%%%%%%%%%%%%%%%%%%%%%%%%%%%%%%%%%%%%%%%%%%%%%%%%%%%%

\textit{Discussion.} Our result~\eqref{E:result} can be viewed either as a higher-dimensional $g$-theorem, or as a generalization of the weak $c$-theorem to include coupling to a higher-dimensional CFT. Indeed, the $g$-theorem itself can be viewed as a monotonicity theorem for a $d=1$ QFT with an RG flow coupled to a $d=2$ CFT. A natural question is whether every monotonicity theorem survives coupling to a higher-dimensional CFT.

Other natural questions arise from further comparisons to existing monotonicity theorems. For example, the strong $c$- and $F$-theorems can be established using strong sub-additivity of EE~\cite{Casini:2004bw,Casini:2006,Casini:2012ei,Casini:2015woa}. Can we establish a strong(est) $b$-theorem, for example using EE? In $d=2$, the $g$-theorem can be violated by a bulk RG flow~\cite{Green:2007wr}. Can a bulk RG flow violate the $b$-theorem? In SUSY theories, $F$- and $a$-maximization provide rigorous tests of the $F$- and $a$-theorems~\cite{Jafferis:2010un,Intriligator:2003jj}. Can $c$-extremization~\cite{Benini:2012cz} be extended to two-dimensional SUSY defects? If so, can that provide tests of the $b$-theorem?

Our result may have implications for many theoretical and experimental systems. One example is a graphene nanoribbon, which at low energy is described by a $d=3$ CFT (free massless Dirac fermions)~\cite{Semenoff:1984dq} on a space with a boundary. Another example is the critical Ising model in $d\geq 3$ with a planar defect, or in $d=3$ with a boundary. Although we assumed parity invariance, our result~\eqref{E:result} is straightforward to generalize to parity-violating theories, and hence may have implications for quantum Hall systems. More abstractly, in string and M-theory, brane intersections can give rise to various DCFTs and BCFTs in $d \geq 3$. What consequences our result may have for all of these systems deserves exploration.

%%%%%%%%%%%%%%%%%%%%%%%%%%%%%%%%%%%%%%%%%%%%%%%%%%%%%%%%%%%%%%%%%%%%%%%%%%%%%%%%%%%%%%%%%%%%%%%

\textit{Acknowledgements.} We are pleased to thank K.~Balasubramanian, A.~Castro, C.~Eling, S.~Hellerman, C.~Herzog, V.~Ker\"anen, D.~Martelli, R.~Myers, D.~Park, E.~Perlmutter, L.~Rastelli, and M.~Taylor for helpful discussions. We also thank M.~Buican, J.~Cardy, J.~Estes, Z.~Komargodski, Y.~Korovin, A.~Schwimmer, K.~Skenderis, A.~Stergiou, and T.~Takayanagi for their helpful comments on the manuscript. K.~J. was supported by the NSF under grant PHY-0969739. A.~O'B. was supported by a University Research Fellowship from the Royal Society of London and a Junior Research Fellowship from Balliol College. We thank the Galileo Galilei Institute for Theoretical Physics for hospitality and the INFN for partial support during the completion of this work.

%%%%%%%%%%%%%%%%%%%%%%%%%%%%%%%%%%%%%%%%%%%%%%%%%%%%%%%%%%%%%%%%%%%%%%%%%%%%%%%%%%%%%%%%%%%%%%%

\appendix

\section{APPENDIX}

%%%%%%%%%%%%%%%%%%%%%%%%%%%%%%%%%%%%%%%%%%%%%%%%%%%%%%%%%%%%%%%%%%%%%%%%%%%%%%%%%%%%%%%%%%%%%%%
\section{A. Some Submanifold Geometry}
%%%%%%%%%%%%%%%%%%%%%%%%%%%%%%%%%%%%%%%%%%%%%%%%%%%%%%%%%%%%%%%%%%%%%%%%%%%%%%%%%%%%%%%%%%%%%%%

\setcounter{equation}{0}
\renewcommand\theequation{A\arabic{equation}}

In this appendix we collect a few standard results from the geometry of submanifolds that will be useful in the subsequent appendices. We consider a manifold of dimension $d$, with coordinates $x^{\mu}$ where $\mu=1,\ldots,d$, and a submanifold of dimension $m<d$, with coordinates $\sigma^a$ where $a=1,\ldots,m$. The submanifold's position is described by embedding functions $X^{\mu}(\sigma^a)$.

The bulk metric $g_{\mu\nu}$ induces a submanifold metric $\hat{g}_{ab}=\partial_a X^{\mu}\partial_b X^{\nu} g_{\mu\nu}(X)$. Using $\hat{g}_{ab}$ and the $X^{\mu}(\sigma^a)$ we define a projector tangential to the submanifold, also called the first fundamental form,
\beq
P^{\mu}_{~\nu} \equiv g_{\nu\rho} \hat{g}^{ab} \partial_a X^{\mu}\partial_b X^{\rho}\,,
\eeq
and a projector normal to the submanifold,
\beq
\label{E:proj}
N^{\mu}_{~\nu} \equiv \delta^{\mu}_{~\nu} - P^{\mu}_{~\nu}\,.
\eeq
The bulk covariant derivative $\nabla_{\mu}$ induces a defect covariant derivative $\hat{\nabla}_a$, which can act on tensors with bulk and defect indices: for a mixed-index tensor $M^{\mu}_a$,
\beq
\hat{\nabla}_a M^{\mu}_b \equiv \partial_a M^{\mu}_b + \Gamma^{\mu}{}_{\nu a} M^{\nu}_b - \hat{\Gamma}^c{}_{ab} M^{\mu}_c\,,
\eeq
with $\Gamma^{\mu}_{~\nu a}$ the pullback of the Levi-Civita connection,
\beq
\Gamma^{\mu}_{~\nu a} \equiv \Gamma^{\mu}{}_{\nu\rho} \partial_a X^{\rho}\,,
\eeq
and with $\hat{\Gamma}^a_{~bc}$ the Levi-Civita connection associated with $\hat{g}_{ab}$. The action of $\hat{\nabla}_a$ on more general tensors follows by the usual rules.

Using $\hat{\nabla}_a$, we define the second fundamental form, $\II^{\mu}{}_{ab} \equiv \hat{\nabla}_a \partial_b X^{\mu}$, which is a normal-vector-valued symmetric tensor: $P^{\mu}_{~\nu} \II^{\nu}_{~ab} = 0$ and $\II^{\mu}_{~ba} = \II^{\mu}_{~ab}$.

In Appendix C we will need the variation of the scalar curvature, $R$, and of the second fundamental form, $\II^{\mu}_{~ab}$, under an infinitesimal Weyl variation, $\delta_{\omega} g_{\mu\nu}=2\omega g_{\mu\nu}$, $\delta_{\omega} X^{\mu}=0$:
\begin{align}
\begin{split}
\delta_{\omega} R &= -2\omega R -2 (d-1) \nabla^2 \omega,
\\
\label{E:deltaRPi}
 \delta_{\omega} \II^{\mu}_{~ab} &= - N^{\mu\nu} \hat{g}_{ab} \partial_{\nu} \omega\,.
\end{split}
\end{align}
The traceless part of the second fundamental form
\beq
\mathring{\II}^{\mu}{}_{ab} \equiv \II^{\mu}{}_{ab} - \frac{1}{m}\,\hat{g}_{ab} \, \hat{g}^{cd}\II^{\mu}{}_{cd}\,,
\eeq
is then Weyl-invariant.

For a submanifold with only one normal direction, let $n^{\mu}$ be the unit-length normal vector field. The extrinsic curvature $K_{ab} \equiv - n_{\mu} \II^{\mu}_{~ab}$, with trace $K \equiv \hat{g}^{ab} K_{ab}$. The Weyl variation of $K$ follows from that of $\II^{\mu}_{~ab}$ in~\eqref{E:deltaRPi},
\beq
\label{E:deltaK}
\delta_{\omega} K = - \omega K + (d-1)n^{\mu}\partial_{\mu}\omega\,.
\eeq
The traceless part of the extrinsic curvature,
\beq
\mathring{K}_{ab} \equiv K_{ab} - \frac{1}{m}\hat{g}_{ab} K,
\eeq
obeys $\mathring{K}_{ab}\mathring{K}^{ab}=\mathring{\II}^{\mu}_{~ab}\mathring{\II}_{\mu}^{~ab}$.

%%%%%%%%%%%%%%%%%%%%%%%%%%%%%%%%%%%%%%%%%%%%%%%%%%%%%%%%%%%%%%%%%%%%%%%%%%%%%%%%%%%%%%%%%%%%%%%
\section{B. Reparameterization Ward Identities}
%%%%%%%%%%%%%%%%%%%%%%%%%%%%%%%%%%%%%%%%%%%%%%%%%%%%%%%%%%%%%%%%%%%%%%%%%%%%%%%%%%%%%%%%%%%%%%%

\setcounter{equation}{0}
\renewcommand\theequation{B\arabic{equation}}

Consider a QFT in dimension $d$ with a defect of dimension $m$, with generating functional of renormalized, connected correlators $W$. The invariance of $W$ under reparameterizations of the bulk coordinates $x^{\mu}$ and defect coordinates $\sigma^a$ leads to Ward identities for the stress tensor, $T^{\mu\nu}$, and displacement operator, $D_{\mu}$. In this appendix we present brief derivations of those Ward identities for the one-point functions $\langle T^{\mu\nu} \rangle$ and $\langle D_{\mu} \rangle$. For more detailed derivations, see for example~\cite{McAvity:1993ue}.

Under an infinitesimal variation of the defect coordinates only, generated by a vector field $\zeta^a$ on the defect,
\beq
\delta_{\zeta} x^{\mu}=0, \quad \delta_{\zeta} \sigma^a = - \zeta^a\,,
\eeq
the bulk metric and embedding functions transform as
\beq
\delta_{\zeta} g_{\mu\nu}=0, \quad \delta_{\zeta} X^{\mu} = \zeta^a \partial_a X^{\mu}\,.
\eeq
Plugging these variations into the $\delta W$ in~\eqref{E:deltaW}, and demanding $\delta_{\zeta} W=0$ for arbitrary $\zeta^a$, we find a Ward identity stating that the components of $\langle D_{\mu} \rangle$ parallel to the defect must vanish:
\beq
\partial_a X^{\mu} \langle D_{\mu} \rangle = 0\,.
\eeq

Under an infinitesimal variation of the bulk coordinates only, generated by a vector field $\chi^{\mu}$ in the bulk,
\beq
\delta_{\chi} x^{\mu}=-\chi^{\mu}, \quad \delta_{\chi} \sigma^a = 0\,,
\eeq
the bulk metric and embedding functions transform as
\beq
\delta_{\chi} g_{\mu\nu} = \nabla_{\mu} \chi_{\nu} + \nabla_{\nu} \chi_{\mu}, \quad \delta_{\chi} X^{\mu}= -\chi^{\mu}\,.
\eeq
Plugging these variations into the $\delta W$ in~\eqref{E:deltaW} and demanding $\delta_{\chi} W=0$ for arbitrary $\chi^{\mu}$ (temporarily ignoring terms involving derivatives of $\delta_{\chi} g_{\mu\nu}$ normal to the defect, which we discuss below), we find a Ward identity for the divergence of $\langle T^{\mu\nu} \rangle$,
\beq
\label{E:reparWard}
\nabla_{\nu}\langle T^{\nu\mu}\rangle = - \delta^{d-m} \langle D^{\mu}\rangle\,,
\eeq
with $\delta^{d-m}$ a Dirac delta function that restricts to the defect. Equivalently, in terms of $\langle T^{\mu\nu}_b \rangle$, $\langle T^{\mu\nu}_d \rangle$, and the normal projector $N^{\mu}_{~\nu}$ in~\eqref{E:proj},~\eqref{E:reparWard} becomes
\beq
\label{E:reparWard2}
\nabla_{\mu}\langle T_b^{\mu\nu} \rangle =-\delta^{d-m} \left(\langle  D^{\nu} \rangle + \hat{\nabla}_a \langle T_d^{a\nu}\rangle \right)- \langle T^{\mu\nu}_d\rangle N_{~\mu}^{\rho}\nabla_{\rho} \delta^{d-m}.
\eeq
If the bulk stress tensor is smooth, then the last term in~\eqref{E:reparWard2} must vanish identically, in which case the defect stress tensor has only parallel components,
\beq
\label{E:reparWard3}
\langle T_d^{\mu\nu}\rangle = \partial_a X^{\mu}\partial_b X^{\nu}\langle T_d^{ab}\rangle\,.
\eeq
If we impose~\eqref{E:reparWard3}, then~\eqref{E:reparWard2} says that energy and momentum parallel to the defect are conserved, but can flow between the bulk and the defect, while momentum normal to the defect is not conserved.

If $\delta_{\chi}W$ includes terms involving normal derivatives of $\delta_{\chi} g_{\mu\nu}$, then $\langle T^{\mu\nu}\rangle$ contains terms proportional to normal derivatives of $\delta^{d-m}$. Nevertheless, the bulk reparameterization Ward identity~\eqref{E:reparWard} is unchanged, although~\eqref{E:reparWard2} will take a more complicated form.

Integrating over the $\delta^{d-m}$ in~\eqref{E:reparWard2}, using a ``Gaussian pillbox'' normal to the defect, gives us Ward identities for the divergence of $\langle T_d^{a\mu}\rangle$ and the normal components of $\langle D_{\mu} \rangle$. For example, for a defect of co-dimension one, or for a boundary, and for the parallel and normal directions, respectively, we find
\begin{align}
\begin{split}
\label{E:wardBdy}
\hat{\nabla}_a \langle T_d^{a\mu}\rangle &= \langle T_b^{\mu\nu}\rangle n_{\nu}\,, 
	\\
 	\langle D_{\mu}\rangle n^{\mu}&= \langle T_b^{\mu\nu}\rangle n_{\mu}n_{\nu}+\langle T_d^{ab}\rangle K_{ab}\,.
\end{split}
\end{align}

%%%%%%%%%%%%%%%%%%%%%%%%%%%%%%%%%%%%%%%%%%%%%%%%%%%%%%%%%%%%%%%%%%%%%%%%%%%%%%%%%%%%%%%%%%%%%%%
\section{C. Free Scalar Boundary Central Charge}
%%%%%%%%%%%%%%%%%%%%%%%%%%%%%%%%%%%%%%%%%%%%%%%%%%%%%%%%%%%%%%%%%%%%%%%%%%%%%%%%%%%%%%%%%%%%%%%

\setcounter{equation}{0}
\renewcommand\theequation{C\arabic{equation}}

In this appendix we consider a free, massless scalar field $\varphi$ on a space with a boundary. First, for $\varphi$ on curved space of any $d$, with conformal coupling to curvature, we determine the allowed BC (linear in $\varphi$), and identify the subset of conformally-invariant BC. Second, for $\varphi$ in $d=3$ flat space with a planar boundary, we compute the boundary central charge $b$ for the two conformally-invariant BC, Dirichlet and Neumann.

Locality, reparameterization invariance, and Weyl invariance fix the form of $\varphi$'s action, $S$, up to a single free parameter: assigning $\varphi$ the Weyl transformation $\varphi \to e^{-\frac{d-2}{2} \omega}\varphi$, we find, using~\eqref{E:deltaRPi} and~\eqref{E:deltaK},
\begin{align}
	\label{E:scalar}
	S= & \frac{1}{2}\int d^dx\sqrt{g} \left\{ (\partial\varphi)^2 + \xi R \varphi^2\right\}
	\\
	\nonumber
	& \quad +\frac{1}{2}\int d^{d-1}\sigma\sqrt{\hat{g}}\left\{  2\xi K \varphi^2 +\lambda \left( \varphi \partial_n \varphi + 2 \xi K \varphi^2\right)\right\},
\end{align}
where $\partial_n \varphi$ denotes $\varphi$'s normal derivative,
\beq
\partial_n \varphi \equiv n^{\mu}\partial_{\mu}\varphi,
\eeq
and $\xi$ and $\lambda$ are dimensionless, real-valued constants. In~\eqref{E:scalar}, Weyl invariance fixes
\beq
\xi = \frac{d-2}{4(d-1)},
\eeq
and also fixes the coefficient of the first boundary term (the term without $\lambda$): that term's Weyl variation must cancel the boundary term produced by the Weyl variation of the bulk terms. However, Weyl invariance does not fix $\lambda$, because the boundary term $\propto \lambda$ is conformally invariant by itself. As a result, $\lambda$ is an exactly marginal boundary coupling.

We can determine all admissible BC by demanding a consistent variational principle for $\varphi$. The variation of $S$ with respect to $\varphi$, evaluated on a solution of the bulk equation of motion, is
\begin{align}
\label{E:variationalPrinciple}
\delta_{\varphi} S =& \frac{1}{2} \int d^{d-1}\sigma \, \sqrt{\hat{g}} \, \left\{  \lambda \, \varphi \, \delta (\partial_n \varphi) \right.
\\
\nonumber
& \qquad \qquad \left . +\left[(2+\lambda) \partial_n\varphi + 4(1+\lambda) \xi K \varphi \right] \delta \varphi \right \}.
\end{align}
The admissible BC are solutions of $\delta_{\varphi} S=0$. For any $\lambda$, admissible BC are the Dirichlet BC $\varphi= 0$, and the Robin BC
\beq
\label{E:robin}
\partial_n \varphi = - 2 \xi K \varphi.
\eeq
In two special cases, additional BC become admissible. If $\lambda=0$, then
\beq
\delta_{\varphi} S =  \int d^{d-1}\sigma \, \sqrt{\hat{g}} \, \left\{ \partial_n\varphi + 2 \xi K \varphi \right\} \delta \varphi,
\eeq
which admits the more general Dirichlet BC that $\varphi$ can be any function of the boundary coordinates $\sigma^a$, while the only admissable Robin BC is~\eqref{E:robin}. If $\lambda = -1$, then
\beq
\delta_{\varphi} S = \frac{1}{2} \int d^{d-1}\sigma \, \sqrt{\hat{g}} \, \left\{ \partial_n \varphi \, \delta \varphi - \varphi \, \delta(\partial_n  \varphi)  \right\},
\eeq
which admits the more general Dirichlet BC that $\varphi$ can be any function of the $\sigma^a$, and the more general Robin BC that $\partial_n \varphi \propto K \varphi$ with any constant proportionality factor. However, for any $\lambda$, including $\lambda=0$ and $\lambda=-1$, the only conformally-invariant BC are the particular Dirichlet BC $\varphi=0$ and the particular Robin BC in~\eqref{E:robin}.

We now restrict to $d=3$ flat space with planar boundary. The two conformally-invariant BC, Dirichlet and Robin, define two free scalar BCFTs. A boundary mass triggers a boundary RG flow from the Robin BCFT to the Dirichlet BCFT. We want to compute $b$ for each of these, to confirm that $b$ decreases along this boundary RG flow. Crucially, $b$ is an intrinsic property of each BCFT, independent of the background manifold, so we can compute $b$ on any convenient manifold. We choose a hemisphere, $\mathbb{HS}^3$, of radius $r$, with metric
\beq
\label{E:hsmetric}
g_{\mathbb{HS}^3} = r^2 (d\theta^2 + \sin^2\theta \, (d\phi^2+\sin^2\phi \,d\psi^2)),
\eeq
where $\theta \in [0,\pi/2]$, with the equator at $\theta=\pi/2$. The Dirichlet BC is $\varphi = 0$ at $\theta=\pi/2$. The equator has $K_{ab}=0$, so the Robin BC~\eqref{E:robin} reduces to a Neumann BC, $\partial_{\theta}\varphi = 0$ at $\theta=\pi/2$. We henceforth refer to the Robin BC as a Neumann BC.

For any BCFT in $d=3$, the $\mathbb{HS}^3$ free energy, $F_{\mathbb{HS}^3}\equiv - \ln Z_{\mathbb{HS}^3}$, is defined only up to a logarithmic ambiguity, due to the Weyl anomaly:
\beq
\label{E:ambiguity1}
F_{\mathbb{HS}^3} = -\frac{\ln (r \mu)}{24\pi}\int d^2\sigma \sqrt{\hat{g}}\left( b \hat{R} + d_1 \mathring{K}_{ab}\mathring{K}^{ab}\right) + \tilde{F},
\eeq
where $\mu$ is an arbitrary energy scale that we must introduce to define the theory, and $\tilde{F}$ is a constant which can depend on exactly marginal couplings. The equator has $\hat{R}=2/r^2$ and $K_{ab}=0$, so
\beq
\label{E:ambiguity2}
F_{\mathbb{HS}^3} = -\frac{b}{3}\ln (r\mu) + \tilde{F}.
\eeq

While $F_{\mathbb{HS}^3}$ is ambiguous, the coefficient of the logarithm, $b/3$, is unambiguous and physical: that coefficient is invariant under a rescaling of $\mu$, and moreover cannot be shifted by a local counterterm. Furthermore, WZ consistency requires $b$ to be independent of any exactly marginal boundary couplings. Alternatively, that independence is a corollary of our $b$-theorem~\cite{Note1}.

In the free scalar BCFT in $d=3$, if we impose either Dirichlet or Neumann BC, then an integration by parts reduces the action $S$ in~\eqref{E:scalar} to a bulk term alone,
\beq
\label{E:scalar2}
S = \frac{1}{2}\int d^3x \sqrt{g} \, \varphi\left( - \nabla^2 + \xi R\right) \varphi,
\eeq
which is manifestly independent of $\lambda$. As a result, $b$ is (trivially) independent of $\lambda$, as expected.

We now compute $F_{\mathbb{HS}^3}$ explicitly, and extract $b$ from the coefficient of $\ln (r\mu)$. From~\eqref{E:scalar2} we find
\beq
\label{E:Fhemi}
F_{\mathbb{HS}^3} =  \frac{1}{2}\ln \det\left( - \nabla^2 + \xi R\right),
\eeq
where the functional determinant is taken over field configurations that respect the BC. Fortunately, we can determine those configurations from the harmonic analysis of $-\nabla^2 + \xi R$ on $\mathbb{S}^3$, as we now review.

The $\mathbb{S}^3$ metric, $g_{\mathbb{S}^3}$, has the same form as $g_{\mathbb{HS}^3}$ in~\eqref{E:hsmetric}, but with $\theta \in [0,\pi]$. Starting now, we switch to units with $r \equiv 1$. The $\mathbb{S}^3$ scalar spherical harmonics, $Y_{jlm}$, transform in the $(\frac{j}{2},\frac{j}{2})$ representation of the $SO(4)$ isometry group of $\mathbb{S}^3$, and by definition obey
\beq
 - \nabla^2 Y_{jlm} = j(j+2)Y_{jlm}.
\eeq
The $Y_{jlm}$ are thus also eigenfunctions of $-\nabla^2 + \xi R$, with eigenvalues (using $\xi R = 3/4$),
\beq
\label{E:eigenvalues}
\Lambda_j \equiv j(j+2) + \frac{3}{4}.
\eeq
In the $Y_{jlm}$, the dependence on $\theta$ and the $\mathbb{S}^2$ coordinates factorizes: the $\theta$ dependence involves only $j$ and $l$, and the dependence on the $\mathbb{S}^2$ coordinates is contained in the $\mathbb{S}^2$ scalar spherical harmonics, $Y_{lm}(\mathbb{S}^2)$,
\beq
\label{E:factorize}
Y_{jlm}(\theta,\mathbb{S}^2) = y_{jl}(\theta) \, Y_{lm}(\mathbb{S}^2).
\eeq

We want to find the eigenfunctions of $-\nabla^2 + \xi R$ that satisfy either Dirichlet or Neumann BC on the $\mathbb{S}^3$ equator, $\theta = \frac{\pi}{2}$. To do so, we exploit the $\mathbb{Z}_2$ symmetry which reflects about the $\mathbb{S}^3$ equator, $\theta \to \pi - \theta$. Let $P$ denote the operator that generates this $\mathbb{Z}_2$ symmetry. The $Y_{jlm}$ furnish an eigenbasis of $P$, with
\beq
PY_{jlm} = \text{sgn}(jl) \, Y_{jlm},
\eeq
where the eigenvalues depend only on $jl$ because $\theta \to \pi - \theta$ acts only on $y_{jl}(\theta)$ in~\eqref{E:factorize}. The $Y_{jlm}$'s with eigenvalue $-1$ under $P$ obey Dirichlet BC on the equator, while the $Y_{jlm}$'s with eigenvalue $+1$ obey Neumann BC. Consequently, on $\mathbb{HS}^3$ the eigenfunctions of $-\nabla^2 + \xi R$ that obey Dirichlet or Neumann BC on the equator are $Y_{jlm}$'s with $jl<0$ or $jl>0$, respectively. These eigenfunctions have degeneracies
\beq
\label{E:degeneracies}
d_j =
	\begin{cases} 
		\frac{1}{2}j(j+1)\,, & (\text{Dirichlet})
		\\ \frac{1}{2}(j+1)(j+2). & (\text{Neumann})
	\end{cases}
\eeq

We now have all the ingredients that we need to compute $b$. From~\eqref{E:Fhemi},~\eqref{E:eigenvalues}, and~\eqref{E:degeneracies} we have
\beq
\label{E:Fhemi2}
F_{\mathbb{HS}^3} = \frac{1}{2}\sum_{j=0}^{\infty} d_j \ln \Lambda_j,
\eeq
which diverges. We thus introduce a heat kernel regulator, with IR cutoff $\epsilon$,
\beq
\ln \Lambda_j = \int_{\epsilon}^{\infty}\frac{dt}{t}e^{-t \Lambda_j},
\eeq
which renders $F_{\mathbb{HS}^3}$ finite, so we can exchange the sum over $j$ with the integration over $t$. Splitting 
\beq
\ln \Lambda_j = \ln \left( j + \frac{1}{2}\right)+\ln \left( j+\frac{3}{2}\right),
\eeq
we find
\beq
F_{\mathbb{HS}^3} = \frac{1}{2}\int_{\epsilon}^{\infty} \frac{dt}{t}\sum_{j=0}^{\infty} d_j \left[ e^{-t \left( j+\frac{1}{2}\right)}+e^{-t\left( j+\frac{3}{2}\right)}\right]\,.
\eeq
We next perform the sum over $j$: with the $d_j$ for Dirichlet BC in~\eqref{E:degeneracies},
\beq
\label{E:jsum}
\sum_{j=0}^{\infty} \frac{1}{2}j(j+1) \, e^{-t j} = \frac{e^{2t}}{(e^t-1)^3},
\eeq
and with the $d_j$ for Neumann BC in~\eqref{E:degeneracies} the sum over $j$ gives $e^t$ times~\eqref{E:jsum}. Integrating in $t$ produces terms $\propto \epsilon^{-3}$, $\epsilon^{-2}$, and $\epsilon^{-1}$, followed by a $\ln(\epsilon)$ term and a constant. Comparing to~\eqref{E:ambiguity2}, we thus identify
\beq
b = \begin{cases} -\frac{1}{16}\,, & (\text{Dirichlet}) \\ \frac{1}{16}\,. & (\text{Neumann}) \\ \end{cases}
\eeq
For the boundary RG flow from Neumann BC to Dirichlet BC, $b_{\textrm{UV}} = 1/16$ and $b_{\textrm{IR}}=-1/16$, confirming that $b_{\textrm{UV}} > b_{\textrm{IR}}$, as mentioned in the main text.

The fact that the $b$'s for Dirichlet and Neumann BC are equal and opposite is not an accident. The degeneracies $d_j$ for Dirichlet and Neumann BC in~\eqref{E:degeneracies} sum to $(j+1)^2$, which is the degeneracy of scalar harmonics on $\mathbb{S}^3$ with angular momentum $j$. From~\eqref{E:Fhemi2}, the sum of the corresponding $F_{\mathbb{HS}^3}$'s is thus $F/2$, with $F$ the $\mathbb{S}^3$ free energy of the conformally coupled scalar. The latter is unambiguous, hence the coefficients of the logarithms in the two $F_{\mathbb{HS}^3}$'s must sum to zero. Moreover, although the constant $\tilde{F}$'s for Dirichlet and Neumann BC are each unphysical, their sum is the physical $F/2$.

The same happens in the free scalar DCFT in $d=3$, where $\varphi$ satisfies either Dirichlet or Neumann BC at a codimension-one defect. If we put these DCFTs on $\mathbb{S}^3$ with the defect at the equator, then the corresponding $F$'s must sum to the $F$ of the free scalar CFT on $\mathbb{S}^3$. In particular, all ambiguities, including the logarithmic ambiguity due to the Weyl anomaly, cancel in the sum over BC. Again, this is not an accident. Consider for example a DCFT with a local Lagrangian, whose partition function is an integral over field configurations obeying conformally-invariant BC at the defect. Summing over those BC, while keeping the location of the defect fixed, amounts to an integral over all field configurations. That suggests a more general conjecture, that in a CFT any observable can be reconstructed from an appropriate sum over conformal defects.

Our result for $b$ in the Dirichlet BCFT, $b=-1/16$, agrees with that of~\cite{Nozaki:2012qd}, but our result for the Neumann BCFT, $b=1/16$, differs from that of~\cite{Nozaki:2012qd}, $b = 7/16$. The $g$-theorem conjectured in~\cite{Nozaki:2012qd} was for the coefficient of the logarithmic term in the free energy on a ball, $\mathbb{B}^3$. That coefficient is equivalent to our $b$ because the boundary of $\mathbb{B}^3$, that is, an $\mathbb{S}^2$, has $\mathring{K}_{ab}=0$. However, a unit-radius $\mathbb{S}^2$ has $K=2$, so the conformally-invariant BC is the Robin BC in~\eqref{E:robin}, not the Neumann BC used in~\cite{Nozaki:2012qd}. The result of~\cite{Nozaki:2012qd} with Neumann BC is therefore not the $b$ of any free scalar BCFT.

\bibliography{gtheorem}

\end{document}